\documentclass[%
 reprint,
 amsmath,amssymb,
 aps,
]{revtex4-2}

\usepackage{graphicx}
\usepackage{dcolumn}
\usepackage{bm}

\begin{document}

\title{First-Principles Calculations on Elasto-optical Properties of $R$Te$_3$}

\author{Kuiqing Tang}
 \affiliation{Department of Physics, University of Oxford}
\date{\today}

\begin{abstract}
Rare-earth tritellurides ($R$Te$_3$) exhibit complex charge-density-wave (CDW) phases intertwined with lattice symmetry, offering a platform to explore unconventional symmetry breaking in correlated materials. Elasto-optical probing, which detects strain-induced changes in birefringence, provides a non-invasive approach to visualize anisotropy and emergent order in these quasi-two-dimensional systems. However, the magnitude and symmetry of the expected optical response remain poorly quantified, hindering experimental interpretation. Here, we perform first-principles calculations of the elastic, dielectric, and piezo-optical tensors of NdTe$_3$ to establish a quantitative framework for strain-induced optical anisotropy. These results establish a quantitative link between lattice strain and optical response in $R$Te$_3$, providing a predictive framework for probing symmetry-breaking states via elasto-birefringence.

\end{abstract}

\maketitle


\section{\label{sec:level1}Introduction}
Rare-earth tritellurides (\(R\mathrm{Te}_3\), where \(R = \mathrm{La} - \mathrm{Yb}\)) are a family of quasi-two-dimensional materials with tunable lattice and electronic properties\cite{Yumigeta2021,Ru2008}. Each compound crystallizes in an orthorhombic \(Cmcm\) structure (space group \(D_{2h}\)), consisting of nearly square-planar Te layers separated by corrugated rare-earth–Te slabs along the \(b\)-axis\cite{DiMasi1995,Yumigeta2021}. The weak van der Waals coupling between layers and the small in-plane anisotropy between the \(a\) and \(c\) axes make these materials highly responsive to external strain, providing a versatile platform for studying strain-tunable electronic order\cite{Hong2022,Siddique2024, He2022}.
\\
\indent A prominent feature of \(R\mathrm{Te}_3\) is the presence of a unidirectional, incommensurate charge-density wave (CDW) arising from Fermi-surface nesting and strong electron–phonon coupling\cite{Laverock2005,Eiter2012,Ru2008}. The primary CDW forms with modulation vector \( \mathbf{q}_c \approx (0,0,2c^*/7) \), and its transition temperature can be tuned from about 500~K in LaTe\(_3\) to 250~K in TmTe\(_3\) by chemical pressure\cite{DiMasi1995,Yumigeta2021}. In heavier rare-earth members (Tb–Tm), a secondary CDW develops at lower temperatures with a perpendicular modulation \( \mathbf{q}_a \approx (5a^*/7,0,0) \)\cite{Ru2008}. These two CDWs can coexist or compete, leading to intricate coupling in the order-parameter space and a strong sensitivity to perturbations such as pressure, impurities, and anisotropic strain\cite{Walmsley2020,Siddique2024,Hong2022}.
\\
\indent Beyond these conventional CDW phases, recent experiments have revealed evidence of unconventional charge order in \(R\mathrm{Te}_3\)\cite{Wang2022}. Using non-linear optical spectroscopy, evidence of a ferroaxial density wave arising from intertwined charge and orbital order has been reported\cite{Singh2025, Wang2022}. These findings indicate that the CDW state in \(R\mathrm{Te}_3\) hosts complex symmetry breaking that cannot be captured by a simple unidirectional modulation, suggesting that external strain may play a key role in stabilizing or revealing these hidden ferroaxial or multipolar orders.
\\
\indent Because such states are strongly tied to lattice anisotropy, non-invasive optical probes capable of detecting subtle symmetry changes are essential\cite{Sacchetti2006,Pfuner2008,Mennel2018}. In particular, birefringence—the polarization-dependent refraction of light—serves as a direct measure of anisotropy in the dielectric tensor and, by extension, of the underlying lattice or electronic symmetry\cite{Narasimhamurty1981,Nye1957}. When stress is applied, additional birefringence is generated through the photoelastic relation expressed in Voigt notation\cite{Nye1957},
\begin{equation}
\Delta B_{\alpha} = p_{\alpha\beta}\, \varepsilon_{\beta},
\end{equation}
which links strain  \(\varepsilon_{\beta}\) to changes in the inverse dielectric tensor \(\Delta B_{\alpha}\) through the piezo-optic tensor \(p_{\alpha\beta}\)\cite{Narasimhamurty1981,Davi2020}. In materials such as $R\mathrm{Te}_3$, where CDW order strongly couples to lattice distortions\cite{Siddique2024,Hong2022}, measuring the strain-induced birefringence (elasto-birefringence) offers a highly sensitive method for detecting subtle or emergent symmetry lowering, including symmetry changes too small to resolve structurally.
\\
\indent Despite the promise of elasto-birefringence as a diagnostic tool, direct experimental measurements in \(R\mathrm{Te}_3\) are nontrivial. The optical setup demands precise strain alignment and control, and the magnitude of the elasto-birefringence signal itself remains uncertain. Without reliable theoretical benchmarks, it is difficult to distinguish genuine symmetry-breaking responses from potential experimental artifacts.
\\
\indent To address these challenges, in this report, we employ first-principles density functional theory (DFT) calculations to determine the elastic moduli, dielectric constants, and photoelastic coefficients of representative \(R\mathrm{Te}_3\) compounds. By analyzing the strain dependence of the dielectric tensor and resulting optical anisotropy, we estimate the coupling between the lattice deformation and optical response. These results not only provide a theoretical foundation for interpreting elasto and optical experiments on \(R\mathrm{Te}_3\) but also shed light on how elasto-birefringence measurement can potentially provide further evidence on unconventional CDW phases in rare-earth tritellurides.
\\
\section{\label{sec:citeref}Methods}
The calculations were performed using density functional theory (DFT) in the Quantum ESPRESSO package\cite{Giannozzi2009,Giannozzi2017}. The initial crystal structure of NdTe$_3$ was taken from the Materials Project database and converted from the CIF file to a Quantum ESPRESSO input format before relaxation\cite{Jain2013,Ong2013}.
The exchange-correlation energy was treated within the generalized gradient approximation (GGA) using the Perdew-Burke-Ernzerhof (PBEsol) functional. The wave functions were expanded in a plane-wave basis set with a kinetic energy cut-off of 544 eV. The core electrons were described using projector augmented wave (PAW) datasets for Nd and ultrasoft pseudopotentials (USPP) for Te. The Brillouin zone integration was carried out over a 10 $\times$ 2 $\times$ 10 Monkhorst-Pack k-point mesh. A cold smearing scheme with a width of 0.02 Ry was employed for the electronic occupations. The lattice constants of the NdTe$_3$ material were optimized until the total energy converged to at least 1.5 mRy.

\section{Results}
\subsection{\label{sec:level2}Elastic Modulus Tensor}
\begin{figure}[t!]
    \centering
    \includegraphics[trim={0cm 0cm 0cm 0cm},clip,width=0.5\textwidth]{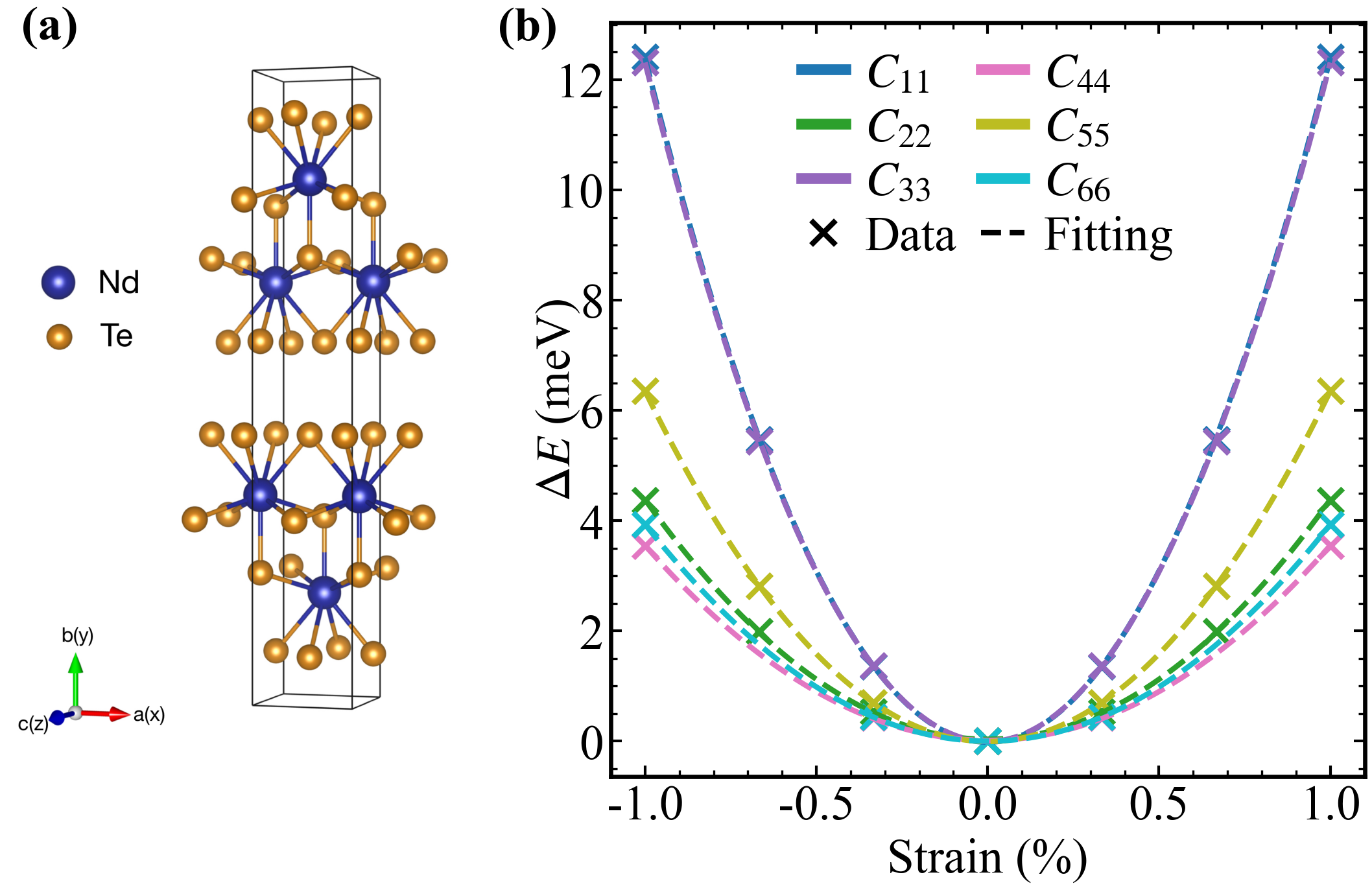}
    \caption{\textbf{Elasto-modulus Calculations of NdTe$_{3}$}. (a) Crystal structure of $R$Te$_{3}$ with specific $R$ to be neodymium. (b) Energy change $\Delta E$ as a function of strain for $\mathrm{NdTe}_3$ under six independent distortions. Cross markers denote first-principles data for the energy difference relative to the ground state, $\Delta E = E(\varepsilon)-E_0$ (Ry) under various external strain, whereas dashed curves represent the quadratic fitting to the data. The elastic constants $C_{11}$, $C_{22}$, $C_{33}$, $C_{44}$, $C_{55}$, and $C_{66}$ are obtained from the zero-strain curvature of the quadratic fits.}
    \label{fig:fig1}
\end{figure}

\begin{table}[t]
\caption{\label{tab:table1}%
Elastic constants of  NdTe$_{3}$, in comparison with literature values of SmTe$_3$\cite{mposti_1313190}}
\begin{ruledtabular}
\begin{tabular}{lcdr}
\textrm{Elastic constant}&
\textrm{NdTe$_3$(GPa)}&
\textrm{SmTe$_3$(GPa)}\cite{mposti_1313190}\\
\colrule
$C_{11}$ & 76.4 & 86\\
$C_{22}$ & 26.8 & 11\\
$C_{33}$ & 75.8 & 92\\
$C_{44}$ & 5.4  & 2\\
$C_{55}$ & 9.8  & 52\\
$C_{66}$ & 6.1  & 2\\
$C_{12}$ & 12.8 & 1\\
$C_{13}$ & 47.0 & 55\\
$C_{23}$ & 20.5 & 3\\
\end{tabular}
\end{ruledtabular}
\end{table}
The elastic properties of NdTe$_3$ were determined from first-principles total-energy calculations under small homogeneous deformations (\(\pm 1\%\)) around the relaxed equilibrium structure. For an orthorhombic crystal, the stiffness tensor contains nine independent components: \(C_{11}\), \(C_{22}\), \(C_{33}\), \(C_{44}\), \(C_{55}\), \(C_{66}\), \(C_{12}\), \(C_{13}\), and \(C_{23}\), expressed in Voigt notation\cite{Nye1957}. Fig.~\ref{fig:fig1} presents representative energy–strain relations for selected deformation modes. The elastic constants were extracted from the energy–strain relation
\begin{equation}
E = E_0 + \frac{1}{2} V_0 \sum_{\alpha,\beta} C_{\alpha\beta} \, \varepsilon_\alpha \varepsilon_\beta,
\end{equation}
where \(E_0\) and \(V_0\) are the equilibrium total energy and cell volume, respectively, and \(\varepsilon_\alpha\) are the strain components in Voigt notation. The resulting symmetrized elastic constants are summarized in Table~\ref{tab:table1}, together with available computational reference values for comparison\cite{mposti_1313190}.

The obtained elastic tensor exhibits strong in-plane stiffness (\(C_{11} \approx C_{33} \gg C_{22}\)), consistent with the quasi-two-dimensional structure of NdTe$_3$, in which covalently bonded Te planes dominate the mechanical response, while the interlayer coupling along the \(b\)-axis is weak due to van der Waals interactions. The relatively small shear constants \(C_{44}\) and \(C_{66}\) further reflect the ease of interlayer sliding and deformation between Te planes. The calculations on the elasto-modulus tensor provide a reliable foundation for evaluating the elasto-optical coefficients in subsequent sections.

\subsection{Dielectric Constants}
\begin{figure}[t!]
    \centering
    \includegraphics[trim={0cm 0cm 0cm 0cm},clip,width=0.48\textwidth]{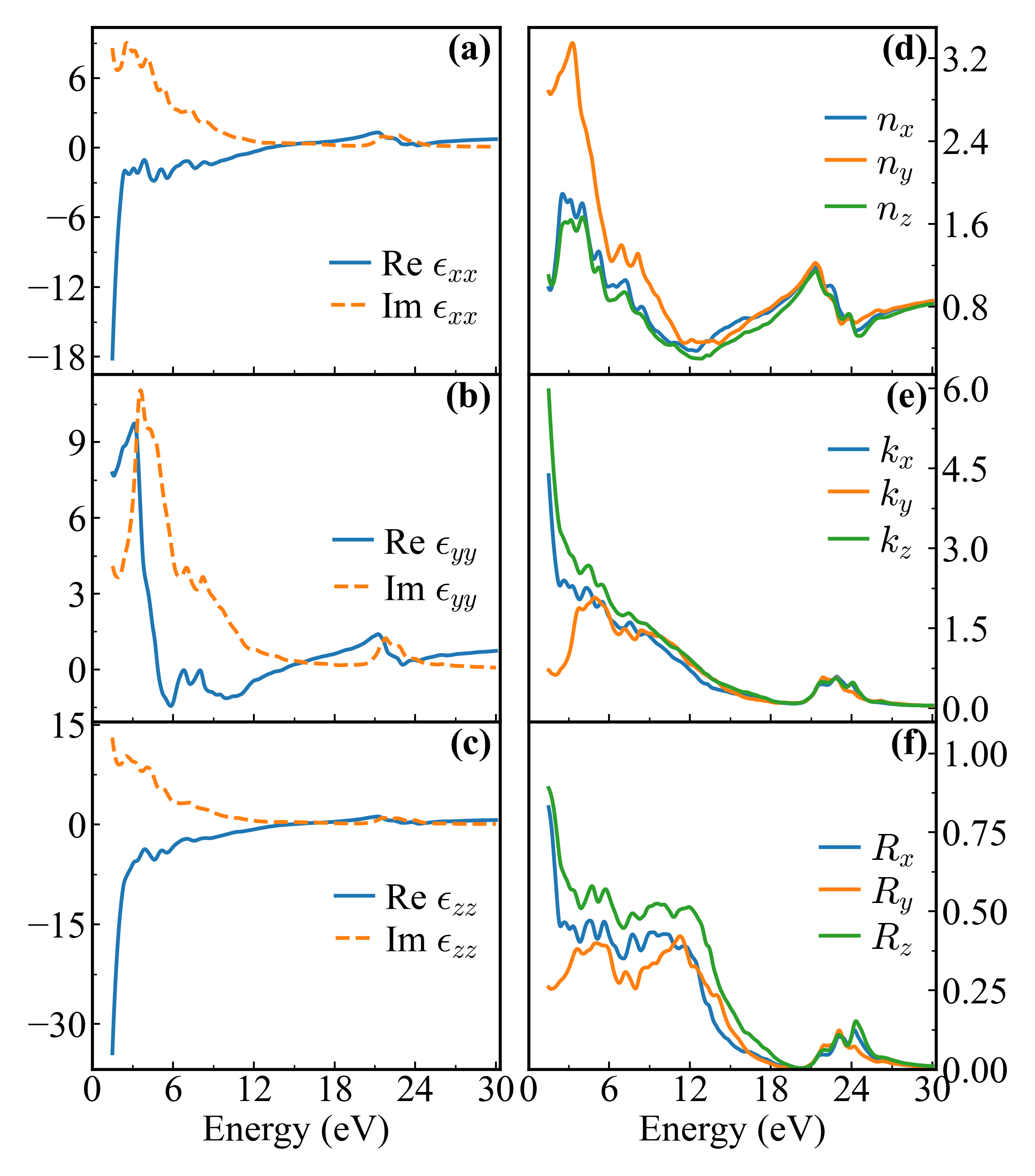}
    \caption{\textbf{Calculations on Optical Properties of NdTe$_{3}$}. (a–c) Real (solid) and imaginary (dashed) parts of the dielectric function
$\epsilon_{ij}$ for $ij=xx,yy,zz$.
(d–e) Real and imaginary parts of the complex refractive index, $n$ and $k$, 
along $x$, $y$, and $z$, obtained from the physical branch of the complex 
refractive index $\tilde{n} = n + i k = \sqrt{\epsilon}$ (chosen such that $k \ge 0$). 
(f) Normal-incidence reflectivity $R$ for $x$, $y$, and $z$, computed as
$R=\left|\tfrac{n-1}{n+1}\right|^{2}$.
The tensor form is constrained by $D_{2h}$ symmetry: in our computational frame
only three components are symmetry-allowed, while $\epsilon_{xy}=\epsilon_{xz}=\epsilon_{yz}=0$.}
    \label{fig:fig2}
\end{figure}
Optical probes offer a powerful, non-invasive means to detect material anisotropy and symmetry breaking on a microscopic scale\cite{Sacchetti2006,Pfuner2008,Mennel2018}. For $R$Te$_3$, where electronic and lattice degrees of freedom are strongly coupled, polarized optical measurements could sensitively reveal strain- or CDW-induced anisotropy\cite{Sacchetti2006,Pfuner2008,Eiter2012}. However, without quantitative simulation, it is difficult to estimate whether the optical contrast will be experimentally measurable—or to distinguish genuine symmetry-breaking signals from artificial effects that may differ by orders of magnitude. To provide this benchmark, we computed the frequency-dependent dielectric tensor $\epsilon(\omega)$ of NdTe$_3$.

Fig.~\ref{fig:fig2} illustrates the complex dielectric function $\epsilon(\omega) = \epsilon_1(\omega) + i \epsilon_2(\omega)$. The real part $\epsilon_1(\omega)$ quantifies the material's polarizability and energy storage capacity, while the imaginary part $\epsilon_2(\omega)$ captures the dissipative processes through interband optical absorption. Both the real and imaginary components display pronounced anisotropy: $\epsilon(\omega)$ varies strongly between in-plane ($\epsilon_{xx}$, $\epsilon_{zz}$) and out-of-plane ($\epsilon_{yy}$) directions. The anisotropic dispersion produces distinct refractive indices and reflectivities along the $a (x)$, $b (y)$, and $c (z)$ axes, confirming the quasi-two-dimensional character of the electronic structure\cite{Sacchetti2006,Pfuner2008}. In particular, above 20~eV, $\epsilon(\omega)$ approaches zero, indicating a plasma frequency near 25~eV and optical transparency beyond the visible range. 




\subsection{Elasto–optical properties: calculation and simulation}
\begin{figure}[t!]
    \centering
    \includegraphics[trim={0cm 0cm 0cm 0cm},clip,width=0.48\textwidth]{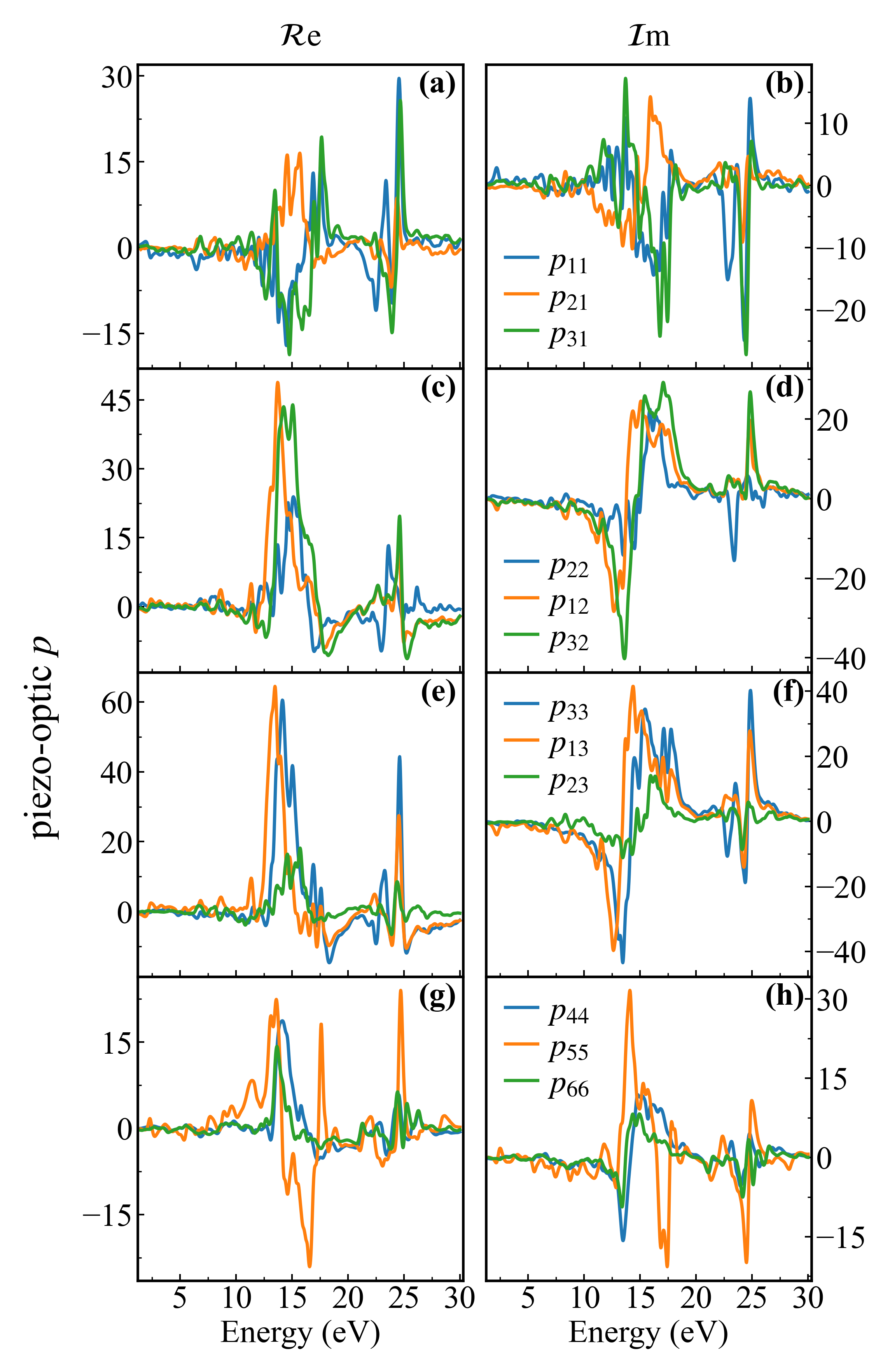}
    \caption{\textbf{Dispersion of piezo-optical tensor of NdTe$_{3}$.}
Real (left column) and imaginary (right column) parts of the selected piezo-optical coefficients $p_{\alpha\beta}$ as functions of photon energy, calculated under $\pm 1\%$ strain.
Each row corresponds to a different applied strain component:
(a,b) $\varepsilon_{1}$-induced response with $p_{11}$, $p_{21}$, $p_{31}$;
(c,d) $\varepsilon_{2}$-induced response with $p_{22}$, $p_{12}$, $p_{32}$;
(e,f) $\varepsilon_{3}$-induced response with $p_{33}$, $p_{13}$, $p_{23}$;
(g,h) shear-strain-induced response with $p_{44}$, $p_{55}$, $p_{66}$, associated with $\varepsilon_{4}$, $\varepsilon_{5}$, $\varepsilon_{6}$, respectively.}
    \label{fig:fig3}
\end{figure}

Strain-induced birefringence provides a sensitive optical probe of symmetry breaking in $R$Te$_3$, motivating a quantitative link between lattice deformation and dielectric response. To this end, we computed the full piezo-optical tensor and used it to simulate the spatial elasto-birefringence patterns expected in a realistic device geometry.

To obtain the energy-dependent piezo-optical coefficients $p_{\alpha\beta}(E)$, we evaluated the inverse dielectric tensor under $\pm 1\%$ uniaxial or shear strain applied to each independent component. For each deformation, the perturbed dielectric tensors $\epsilon_{ij}(E,\pm \Delta \varepsilon_\beta)$ were computed, converted to impermeability form $B(E, \pm \Delta \varepsilon_\beta)$, and combined through a central finite difference to yield $p_{\alpha\beta}(E)$ (Fig.~\ref{fig:fig3}). In $D_{2h}$ symmetry, twelve components are symmetry-allowed, and the calculated spectra exhibit pronounced anisotropy. Representative values at HeNe wavelength 632~nm ($E=1.96$~eV) are listed in Table~II: the normal-stress piezo-optical coefficients dominate, while shear terms such as $p_{44}, p_{55}, p_{66}$ are smaller.

The elasto-birefringence follows directly from the photoelastic relation
\begin{equation}
\Delta B_\alpha(\mathbf{r}) = p_{\alpha\beta} \, \varepsilon_\beta(\mathbf{r}),
\end{equation}
which perturbs the impermeability tensor as
\begin{equation}
B(\mathbf{r}) = B_0 + \Delta B(\mathbf{r}).
\end{equation}
For small perturbations, the corresponding changes in the principal refractive indices satisfy
\begin{equation}
\Delta n_i(\mathbf{r}) \simeq -\frac{1}{2} n_i^3 \, \Delta B_{ii}(\mathbf{r}),
\end{equation}
from which the spatial birefringence
\begin{equation}
\delta \Delta n(\mathbf{r}) = \Delta n_{\rm fast} - \Delta n_{\rm slow}
\end{equation}
is obtained.

\begin{figure}[h]
    \centering
    \includegraphics[width=\linewidth]{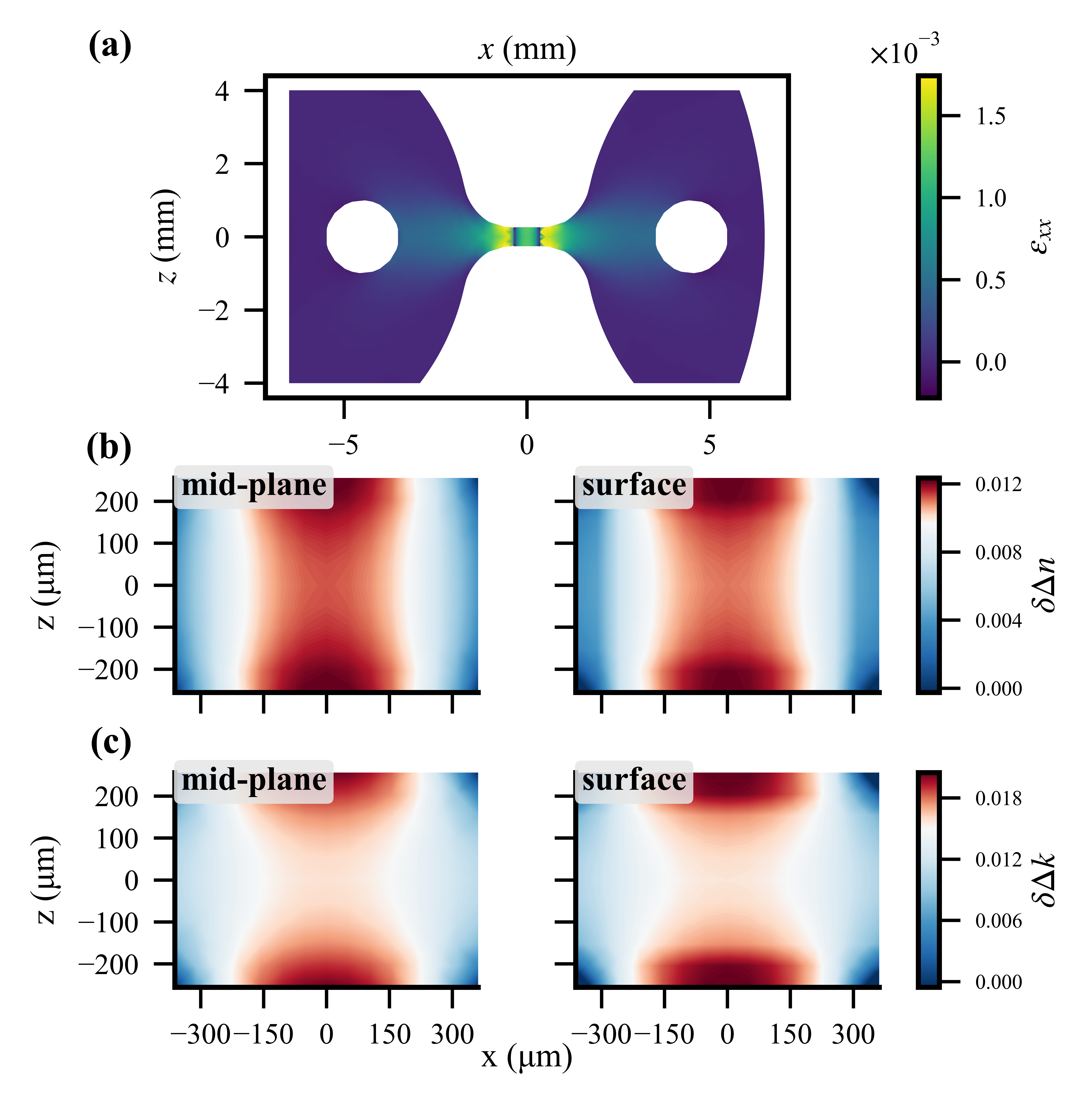}
   \caption{\textbf{Normal strain $\varepsilon_{xx}$ and elasto-birefringence.} (a) Top-view distribution of the normal strain component $\varepsilon_{xx}$ in the bow-tie plate. The strain field is obtained from a finite-element simulation in which one circular hole is fixed while a prescribed in-plane displacement is applied to the opposite hole. The center rectangle lies NdTe$_3$ material. (b,c) Spatial distribution of the elasto-birefringence at photon energy $E \simeq 1.96~\mathrm{eV}$. The elasto-optic response is evaluated from the complex dielectric tensor $\epsilon(\omega)$ and the piezo-optic tensor $p_{\alpha\beta}(\omega)$, combined with the full three-dimensional strain field from finite-element simulations. (b) and (c) are respectively real and imaginary part of the elasto-birefringence $\delta\Delta n + i\delta\Delta k = \big[(\tilde{n}_x - \tilde{n}_z) - (\tilde{n}^0_x - \tilde{n}^0_z)\big]$. Here, ``surface'' indicates the top interface of the material and ``mid-plane'' indicates the middle cross-section of the material between the bottom interface and the top surface.}

    \label{fig:configuration}
\end{figure}

To model experimental conditions, we considered an $R$Te$_3$ flake bonded to a Ti bow-tie substrate (Fig.~\ref{fig:configuration}). The mechanically compliant $R$Te$_3$ layer requires a rigid support to prevent buckling and to impose a well-defined in-plane strain. Finite-element simulations of this device show a predominantly uniaxial strain field concentrated near the narrow waist of the bow-tie, with two symmetric high-strain edge regions along the loading direction.

Combining the simulated strain field with the piezo-optical tensor yields the spatial distribution of the complex elasto-birefringence at 1.96~eV (Fig.~\ref{fig:configuration}). Both surface and mid-plane cross-sections show a relatively uniform elasto-birefringence response in the central region of the sample, with a magnitude of $\sim 1\%$, making it accessible to experimental measurement.

\begin{table}[t]
\caption{\label{tab:table2}%
Complex piezo\mbox{-}optical tensor components $p_{\alpha\beta}$ of NdTe$_3$ at
a photon energy of $E=1.96$~eV. Both real and imaginary parts are listed
according to $p_{\alpha\beta}=p_{\alpha\beta}^\prime + i\,p_{\alpha\beta}^{\prime\prime}$.
}
\begin{ruledtabular}
\begin{tabular}{lcc}
\textrm{Piezo-optical tensor index} & \textrm{Complex value} \\
\colrule
$p_{11}$ & $1.06\,+\,1.57\,i$ \\
$p_{21}$ & $-0.13\,-\,0.41\,i$   \\
$p_{31}$ & $0.38\,+\,0.36\,i$  \\
$p_{22}$ & $0.22\,+\,0.70\,i$  \\
$p_{12}$ & $-1.17\,-\,1.46\,i$  \\
$p_{32}$ & $-0.78\,-\,0.56\,i$  \\
$p_{33}$ & $-0.46\,-\,0.56\,i$  \\
$p_{13}$ & $-1.36\,-\,2.59\,i$   \\
$p_{23}$ & $0.08\,-\,0.50\,i$   \\
$p_{44}$ & $0.19\,-\,0.31\,i$  \\
$p_{55}$ & $-0.25\,-\,0.52\,i$   \\
$p_{66}$ & $-0.20\,-\,0.23\,i$  \\

\end{tabular}
\end{ruledtabular}
\end{table}




\subsection{Symmetry analysis}
Unconventional charge-density-wave (CDW) states in $R$Te$_3$ have been proposed to host additional symmetry-breaking beyond the orthorhombic \(D_{2h}\) (space group Cmcm) lattice — for example, ferroaxial or monoclinic order that reduces the point group\cite{Singh2025,Wang2022,Eiter2012}. Such hidden order may not produce a large change in conventional transport, but it imposes new selection rules on optical and elasto-optical tensors; this makes piezo-optical measurements a sensitive probe of additional symmetry lowering associated with an unconventional CDW transition\cite{Mennel2018,Stoehr2020, singh2024}.

Group theory constrains the piezo-optic (photoelastic) tensor \(p_{ijkl}\) in the parent \(D_{2h}\) symmetry to a block-diagonal form in Voigt notation (indices \(\alpha,\beta=1\ldots6\)):
\[
p^{(D_{2h})} =
\begin{pmatrix}
p_{11} & p_{12} & p_{13} & 0 & 0 & 0\\
p_{21} & p_{22} & p_{23} & 0 & 0 & 0\\
p_{31} & p_{32} & p_{33} & 0 & 0 & 0\\
0 & 0 & 0 & p_{44} & 0 & 0\\
0 & 0 & 0 & 0 & p_{55} & 0\\
0 & 0 & 0 & 0 & 0 & p_{66}
\end{pmatrix}.
\]

Crucially, cross-couplings between the normal-strain block \((1,2,3)\) and shear block \((4,5,6)\) are symmetry-forbidden in \(D_{2h}\)\cite{Nye1957,Narasimhamurty1981,Powell2010}. As a result, certain experimentally measurable optical channels (e.g. in-plane shear driving out-of-plane birefringence) must vanish by symmetry for an unbroken \(D_{2h}\) crystal.

If the system undergoes an additional transition that lowers the symmetry from orthorhombic \(D_{2h}\) to monoclinic \(C_{2h}\) (2/m) with the unique two-fold axis aligned along the interlayer \(y\)–direction, previously forbidden piezo–optic tensor components become finite \cite{Nye1957,Powell2010,Narasimhamurty1981}. In this setting, the most general symmetry–allowed Voigt form of the piezo–optic tensor reads
\[
p^{(C_{2h})} =
\begin{pmatrix}
p_{11} & p_{12} & p_{13} & 0      & p_{15} & 0 \\
p_{21} & p_{22} & p_{23} & 0      & p_{25} & 0 \\
p_{31} & p_{32} & p_{33} & 0      & p_{35} & 0 \\
0      & 0      & 0      & p_{44} & 0      & p_{46} \\
p_{51} & p_{52} & p_{53} & 0      & p_{55} & 0 \\
0      & 0      & 0      & p_{64} & 0      & p_{66}
\end{pmatrix}.
\]

The off-diagonal elements $p_{15}$, $p_{25}$, $p_{35}$, and $p_{46}$ can thus serve as order parameters for probing a potential spontaneous symmetry-breaking state in $R$Te$_3$.

\bibliography{RTe3}

\begin{thebibliography}{25}%
\makeatletter
\providecommand \@ifxundefined [1]{%
 \@ifx{#1\undefined}
}%
\providecommand \@ifnum [1]{%
 \ifnum #1\expandafter \@firstoftwo
 \else \expandafter \@secondoftwo
 \fi
}%
\providecommand \@ifx [1]{%
 \ifx #1\expandafter \@firstoftwo
 \else \expandafter \@secondoftwo
 \fi
}%
\providecommand \natexlab [1]{#1}%
\providecommand \enquote  [1]{``#1''}%
\providecommand \bibnamefont  [1]{#1}%
\providecommand \bibfnamefont [1]{#1}%
\providecommand \citenamefont [1]{#1}%
\providecommand \href@noop [0]{\@secondoftwo}%
\providecommand \href [0]{\begingroup \@sanitize@url \@href}%
\providecommand \@href[1]{\@@startlink{#1}\@@href}%
\providecommand \@@href[1]{\endgroup#1\@@endlink}%
\providecommand \@sanitize@url [0]{\catcode `\\12\catcode `\$12\catcode `\&12\catcode `\#12\catcode `\^12\catcode `\_12\catcode `\%12\relax}%
\providecommand \@@startlink[1]{}%
\providecommand \@@endlink[0]{}%
\providecommand \url  [0]{\begingroup\@sanitize@url \@url }%
\providecommand \@url [1]{\endgroup\@href {#1}{\urlprefix }}%
\providecommand \urlprefix  [0]{URL }%
\providecommand \Eprint [0]{\href }%
\providecommand \doibase [0]{https://doi.org/}%
\providecommand \selectlanguage [0]{\@gobble}%
\providecommand \bibinfo  [0]{\@secondoftwo}%
\providecommand \bibfield  [0]{\@secondoftwo}%
\providecommand \translation [1]{[#1]}%
\providecommand \BibitemOpen [0]{}%
\providecommand \bibitemStop [0]{}%
\providecommand \bibitemNoStop [0]{.\EOS\space}%
\providecommand \EOS [0]{\spacefactor3000\relax}%
\providecommand \BibitemShut  [1]{\csname bibitem#1\endcsname}%
\let\auto@bib@innerbib\@empty
\bibitem [{\citenamefont {Yumigeta}\ \emph {et~al.}(2021)\citenamefont {Yumigeta}, \citenamefont {Qin}, \citenamefont {Li}, \citenamefont {Blei}, \citenamefont {Attarde}, \citenamefont {Kopas},\ and\ \citenamefont {Tongay}}]{Yumigeta2021}%
  \BibitemOpen
  \bibfield  {author} {\bibinfo {author} {\bibfnamefont {K.}~\bibnamefont {Yumigeta}}, \bibinfo {author} {\bibfnamefont {Y.}~\bibnamefont {Qin}}, \bibinfo {author} {\bibfnamefont {H.}~\bibnamefont {Li}}, \bibinfo {author} {\bibfnamefont {M.}~\bibnamefont {Blei}}, \bibinfo {author} {\bibfnamefont {Y.}~\bibnamefont {Attarde}}, \bibinfo {author} {\bibfnamefont {C.}~\bibnamefont {Kopas}},\ and\ \bibinfo {author} {\bibfnamefont {S.}~\bibnamefont {Tongay}},\ }\bibfield  {title} {\bibinfo {title} {Advances in rare-earth tritelluride quantum materials: Structure, properties, and synthesis},\ }\href {https://doi.org/10.1002/advs.202004762} {\bibfield  {journal} {\bibinfo  {journal} {Advanced Science}\ }\textbf {\bibinfo {volume} {8}},\ \bibinfo {pages} {2004762} (\bibinfo {year} {2021})}\BibitemShut {NoStop}%
\bibitem [{\citenamefont {Ru}(2008)}]{Ru2008}%
  \BibitemOpen
  \bibfield  {author} {\bibinfo {author} {\bibfnamefont {N.}~\bibnamefont {Ru}},\ }\emph {\bibinfo {title} {Charge Density Wave Formation in Rare-Earth Tritellurides}},\ \href {https://ui.adsabs.harvard.edu/abs/2008PhDT........31R} {\bibinfo {type} {Phd thesis}},\ \bibinfo  {school} {Stanford University} (\bibinfo {year} {2008})\BibitemShut {NoStop}%
\bibitem [{\citenamefont {DiMasi}\ \emph {et~al.}(1995)\citenamefont {DiMasi}, \citenamefont {Aronson}, \citenamefont {Mansfield}, \citenamefont {Foran},\ and\ \citenamefont {Lee}}]{DiMasi1995}%
  \BibitemOpen
  \bibfield  {author} {\bibinfo {author} {\bibfnamefont {E.}~\bibnamefont {DiMasi}}, \bibinfo {author} {\bibfnamefont {M.~C.}\ \bibnamefont {Aronson}}, \bibinfo {author} {\bibfnamefont {J.~F.}\ \bibnamefont {Mansfield}}, \bibinfo {author} {\bibfnamefont {B.}~\bibnamefont {Foran}},\ and\ \bibinfo {author} {\bibfnamefont {S.}~\bibnamefont {Lee}},\ }\bibfield  {title} {\bibinfo {title} {Chemical pressure and charge-density waves in rare-earth tritellurides},\ }\href {https://doi.org/10.1103/PhysRevB.52.14516} {\bibfield  {journal} {\bibinfo  {journal} {Phys. Rev. B}\ }\textbf {\bibinfo {volume} {52}},\ \bibinfo {pages} {14516} (\bibinfo {year} {1995})}\BibitemShut {NoStop}%
\bibitem [{\citenamefont {Hong}\ \emph {et~al.}(2022)\citenamefont {Hong}, \citenamefont {Wei}, \citenamefont {Liang},\ and\ \citenamefont {Lu}}]{Hong2022}%
  \BibitemOpen
  \bibfield  {author} {\bibinfo {author} {\bibfnamefont {Y.}~\bibnamefont {Hong}}, \bibinfo {author} {\bibfnamefont {Q.}~\bibnamefont {Wei}}, \bibinfo {author} {\bibfnamefont {X.}~\bibnamefont {Liang}},\ and\ \bibinfo {author} {\bibfnamefont {W.}~\bibnamefont {Lu}},\ }\bibfield  {title} {\bibinfo {title} {Origin and strain tuning of charge density wave in {LaTe$_3$}},\ }\href {https://doi.org/10.1016/j.physb.2022.413988} {\bibfield  {journal} {\bibinfo  {journal} {Physica B: Condensed Matter}\ }\textbf {\bibinfo {volume} {639}},\ \bibinfo {pages} {413988} (\bibinfo {year} {2022})}\BibitemShut {NoStop}%
\bibitem [{\citenamefont {Siddique}\ \emph {et~al.}(2024)\citenamefont {Siddique}, \citenamefont {Hart}, \citenamefont {Niedzielski}, \citenamefont {Singha}, \citenamefont {Han}, \citenamefont {Funni}, \citenamefont {Colletta}, \citenamefont {Kiani}, \citenamefont {Schnitzer}, \citenamefont {Williams}, \citenamefont {Kourkoutis}, \citenamefont {Zhu}, \citenamefont {Schoop}, \citenamefont {Arias},\ and\ \citenamefont {Cha}}]{Siddique2024}%
  \BibitemOpen
  \bibfield  {author} {\bibinfo {author} {\bibfnamefont {S.}~\bibnamefont {Siddique}}, \bibinfo {author} {\bibfnamefont {J.~L.}\ \bibnamefont {Hart}}, \bibinfo {author} {\bibfnamefont {D.}~\bibnamefont {Niedzielski}}, \bibinfo {author} {\bibfnamefont {R.}~\bibnamefont {Singha}}, \bibinfo {author} {\bibfnamefont {M.-G.}\ \bibnamefont {Han}}, \bibinfo {author} {\bibfnamefont {S.~D.}\ \bibnamefont {Funni}}, \bibinfo {author} {\bibfnamefont {M.}~\bibnamefont {Colletta}}, \bibinfo {author} {\bibfnamefont {M.~T.}\ \bibnamefont {Kiani}}, \bibinfo {author} {\bibfnamefont {N.}~\bibnamefont {Schnitzer}}, \bibinfo {author} {\bibfnamefont {N.~L.}\ \bibnamefont {Williams}}, \bibinfo {author} {\bibfnamefont {L.~F.}\ \bibnamefont {Kourkoutis}}, \bibinfo {author} {\bibfnamefont {Y.}~\bibnamefont {Zhu}}, \bibinfo {author} {\bibfnamefont {L.~M.}\ \bibnamefont {Schoop}}, \bibinfo {author} {\bibfnamefont {T.~A.}\ \bibnamefont {Arias}},\ and\ \bibinfo {author} {\bibfnamefont {J.~J.}\ \bibnamefont {Cha}},\ }\bibfield  {title}
  {\bibinfo {title} {Realignment and suppression of charge density waves in the rare-earth tritellurides ${R\mathrm{Te}}_{3}$ $({R}=\mathrm{La}, \mathrm{Gd}, \mathrm{Er})$},\ }\href {https://doi.org/10.1103/PhysRevB.110.014111} {\bibfield  {journal} {\bibinfo  {journal} {Phys. Rev. B}\ }\textbf {\bibinfo {volume} {110}},\ \bibinfo {pages} {014111} (\bibinfo {year} {2024})}\BibitemShut {NoStop}%
\bibitem [{\citenamefont {He}\ \emph {et~al.}(2022)\citenamefont {He}, \citenamefont {Wang}, \citenamefont {Du}, \citenamefont {Qin},\ and\ \citenamefont {Qiu}}]{He2022}%
  \BibitemOpen
  \bibfield  {author} {\bibinfo {author} {\bibfnamefont {Q.}~\bibnamefont {He}}, \bibinfo {author} {\bibfnamefont {M.}~\bibnamefont {Wang}}, \bibinfo {author} {\bibfnamefont {Y.}~\bibnamefont {Du}}, \bibinfo {author} {\bibfnamefont {Q.}~\bibnamefont {Qin}},\ and\ \bibinfo {author} {\bibfnamefont {W.}~\bibnamefont {Qiu}},\ }\bibfield  {title} {\bibinfo {title} {Quantitative characterization of the anisotropy of the stress-optical properties of polyethylene terephthalate films based on the photoelastic method},\ }\href {https://doi.org/10.3390/polym14163257} {\bibfield  {journal} {\bibinfo  {journal} {Polymers}\ }\textbf {\bibinfo {volume} {14}},\ \bibinfo {pages} {3257} (\bibinfo {year} {2022})}\BibitemShut {NoStop}%
\bibitem [{\citenamefont {Laverock}\ \emph {et~al.}(2005)\citenamefont {Laverock}, \citenamefont {Dugdale}, \citenamefont {Major}, \citenamefont {Alam}, \citenamefont {Ru}, \citenamefont {Fisher}, \citenamefont {Santi},\ and\ \citenamefont {Bruno}}]{Laverock2005}%
  \BibitemOpen
  \bibfield  {author} {\bibinfo {author} {\bibfnamefont {J.}~\bibnamefont {Laverock}}, \bibinfo {author} {\bibfnamefont {S.~B.}\ \bibnamefont {Dugdale}}, \bibinfo {author} {\bibfnamefont {Z.}~\bibnamefont {Major}}, \bibinfo {author} {\bibfnamefont {M.~A.}\ \bibnamefont {Alam}}, \bibinfo {author} {\bibfnamefont {N.}~\bibnamefont {Ru}}, \bibinfo {author} {\bibfnamefont {I.~R.}\ \bibnamefont {Fisher}}, \bibinfo {author} {\bibfnamefont {G.}~\bibnamefont {Santi}},\ and\ \bibinfo {author} {\bibfnamefont {E.}~\bibnamefont {Bruno}},\ }\bibfield  {title} {\bibinfo {title} {Fermi surface nesting and charge-density wave formation in rare-earth tritellurides},\ }\href {https://doi.org/10.1103/PhysRevB.71.085114} {\bibfield  {journal} {\bibinfo  {journal} {Phys. Rev. B}\ }\textbf {\bibinfo {volume} {71}},\ \bibinfo {pages} {085114} (\bibinfo {year} {2005})}\BibitemShut {NoStop}%
\bibitem [{\citenamefont {Eiter}\ \emph {et~al.}(2013)\citenamefont {Eiter}, \citenamefont {Lavagnini}, \citenamefont {Hackl}, \citenamefont {Nowadnick}, \citenamefont {Kemper}, \citenamefont {Devereaux}, \citenamefont {Chu}, \citenamefont {Analytis}, \citenamefont {Fisher},\ and\ \citenamefont {Degiorgi}}]{Eiter2012}%
  \BibitemOpen
  \bibfield  {author} {\bibinfo {author} {\bibfnamefont {H.-M.}\ \bibnamefont {Eiter}}, \bibinfo {author} {\bibfnamefont {M.}~\bibnamefont {Lavagnini}}, \bibinfo {author} {\bibfnamefont {R.}~\bibnamefont {Hackl}}, \bibinfo {author} {\bibfnamefont {E.~A.}\ \bibnamefont {Nowadnick}}, \bibinfo {author} {\bibfnamefont {A.~F.}\ \bibnamefont {Kemper}}, \bibinfo {author} {\bibfnamefont {T.~P.}\ \bibnamefont {Devereaux}}, \bibinfo {author} {\bibfnamefont {J.-H.}\ \bibnamefont {Chu}}, \bibinfo {author} {\bibfnamefont {J.~G.}\ \bibnamefont {Analytis}}, \bibinfo {author} {\bibfnamefont {I.~R.}\ \bibnamefont {Fisher}},\ and\ \bibinfo {author} {\bibfnamefont {L.}~\bibnamefont {Degiorgi}},\ }\bibfield  {title} {\bibinfo {title} {Alternative route to charge density wave formation in multiband systems},\ }\href {https://doi.org/10.1073/pnas.1214745110} {\bibfield  {journal} {\bibinfo  {journal} {Proceedings of the National Academy of Sciences}\ }\textbf {\bibinfo {volume} {110}},\ \bibinfo {pages} {64} (\bibinfo {year}
  {2013})}\BibitemShut {NoStop}%
\bibitem [{\citenamefont {Walmsley}\ \emph {et~al.}(2020)\citenamefont {Walmsley}, \citenamefont {Aeschlimann}, \citenamefont {Straquadine}, \citenamefont {Giraldo-Gallo}, \citenamefont {Riggs}, \citenamefont {Chan}, \citenamefont {McDonald},\ and\ \citenamefont {Fisher}}]{Walmsley2020}%
  \BibitemOpen
  \bibfield  {author} {\bibinfo {author} {\bibfnamefont {P.}~\bibnamefont {Walmsley}}, \bibinfo {author} {\bibfnamefont {S.}~\bibnamefont {Aeschlimann}}, \bibinfo {author} {\bibfnamefont {J.~A.~W.}\ \bibnamefont {Straquadine}}, \bibinfo {author} {\bibfnamefont {P.}~\bibnamefont {Giraldo-Gallo}}, \bibinfo {author} {\bibfnamefont {S.~C.}\ \bibnamefont {Riggs}}, \bibinfo {author} {\bibfnamefont {M.~K.}\ \bibnamefont {Chan}}, \bibinfo {author} {\bibfnamefont {R.~D.}\ \bibnamefont {McDonald}},\ and\ \bibinfo {author} {\bibfnamefont {I.~R.}\ \bibnamefont {Fisher}},\ }\bibfield  {title} {\bibinfo {title} {Magnetic breakdown and charge density wave formation: A quantum oscillation study of the rare-earth tritellurides},\ }\href {https://doi.org/10.1103/PhysRevB.102.045150} {\bibfield  {journal} {\bibinfo  {journal} {Phys. Rev. B}\ }\textbf {\bibinfo {volume} {102}},\ \bibinfo {pages} {045150} (\bibinfo {year} {2020})}\BibitemShut {NoStop}%
\bibitem [{\citenamefont {Wang}\ \emph {et~al.}(2022)\citenamefont {Wang}, \citenamefont {Petrides}, \citenamefont {McNamara}, \citenamefont {Hosen}, \citenamefont {Lei}, \citenamefont {Wu}, \citenamefont {Hart}, \citenamefont {Lv}, \citenamefont {Yan}, \citenamefont {Xiao}, \citenamefont {Cha}, \citenamefont {Narang}, \citenamefont {Schoop},\ and\ \citenamefont {Burch}}]{Wang2022}%
  \BibitemOpen
  \bibfield  {author} {\bibinfo {author} {\bibfnamefont {Y.}~\bibnamefont {Wang}}, \bibinfo {author} {\bibfnamefont {I.}~\bibnamefont {Petrides}}, \bibinfo {author} {\bibfnamefont {G.}~\bibnamefont {McNamara}}, \bibinfo {author} {\bibfnamefont {M.~M.}\ \bibnamefont {Hosen}}, \bibinfo {author} {\bibfnamefont {S.}~\bibnamefont {Lei}}, \bibinfo {author} {\bibfnamefont {Y.-C.}\ \bibnamefont {Wu}}, \bibinfo {author} {\bibfnamefont {J.~L.}\ \bibnamefont {Hart}}, \bibinfo {author} {\bibfnamefont {H.}~\bibnamefont {Lv}}, \bibinfo {author} {\bibfnamefont {J.}~\bibnamefont {Yan}}, \bibinfo {author} {\bibfnamefont {D.}~\bibnamefont {Xiao}}, \bibinfo {author} {\bibfnamefont {J.~J.}\ \bibnamefont {Cha}}, \bibinfo {author} {\bibfnamefont {P.}~\bibnamefont {Narang}}, \bibinfo {author} {\bibfnamefont {L.~M.}\ \bibnamefont {Schoop}},\ and\ \bibinfo {author} {\bibfnamefont {K.~S.}\ \bibnamefont {Burch}},\ }\bibfield  {title} {\bibinfo {title} {Axial {Higgs} mode detected by quantum pathway interference in {$R$Te$_3$}},\ }\href
  {https://doi.org/10.1038/s41586-022-04746-6} {\bibfield  {journal} {\bibinfo  {journal} {Nature}\ }\textbf {\bibinfo {volume} {606}},\ \bibinfo {pages} {896} (\bibinfo {year} {2022})}\BibitemShut {NoStop}%
\bibitem [{\citenamefont {Singh}\ \emph {et~al.}(2025)\citenamefont {Singh}, \citenamefont {McNamara}, \citenamefont {Kim}, \citenamefont {Siddique}, \citenamefont {Funni}, \citenamefont {Zhang}, \citenamefont {Luo}, \citenamefont {Sakrikar}, \citenamefont {Kenney}, \citenamefont {Singha}, \citenamefont {Alekseev}, \citenamefont {Ghorashi}, \citenamefont {Hicken}, \citenamefont {Baines}, \citenamefont {Luetkens}, \citenamefont {Wang}, \citenamefont {Plisson}, \citenamefont {Geiwitz}, \citenamefont {Occhialini}, \citenamefont {Comin}, \citenamefont {Graf}, \citenamefont {Zhao}, \citenamefont {Cano}, \citenamefont {Fernandes}, \citenamefont {Cha}, \citenamefont {Schoop},\ and\ \citenamefont {Burch}}]{Singh2025}%
  \BibitemOpen
  \bibfield  {author} {\bibinfo {author} {\bibfnamefont {B.}~\bibnamefont {Singh}}, \bibinfo {author} {\bibfnamefont {G.}~\bibnamefont {McNamara}}, \bibinfo {author} {\bibfnamefont {K.-M.}\ \bibnamefont {Kim}}, \bibinfo {author} {\bibfnamefont {S.}~\bibnamefont {Siddique}}, \bibinfo {author} {\bibfnamefont {S.~D.}\ \bibnamefont {Funni}}, \bibinfo {author} {\bibfnamefont {W.}~\bibnamefont {Zhang}}, \bibinfo {author} {\bibfnamefont {X.}~\bibnamefont {Luo}}, \bibinfo {author} {\bibfnamefont {P.}~\bibnamefont {Sakrikar}}, \bibinfo {author} {\bibfnamefont {E.~M.}\ \bibnamefont {Kenney}}, \bibinfo {author} {\bibfnamefont {R.}~\bibnamefont {Singha}}, \bibinfo {author} {\bibfnamefont {S.}~\bibnamefont {Alekseev}}, \bibinfo {author} {\bibfnamefont {S.~A.~A.}\ \bibnamefont {Ghorashi}}, \bibinfo {author} {\bibfnamefont {T.~J.}\ \bibnamefont {Hicken}}, \bibinfo {author} {\bibfnamefont {C.}~\bibnamefont {Baines}}, \bibinfo {author} {\bibfnamefont {H.}~\bibnamefont {Luetkens}}, \bibinfo {author} {\bibfnamefont
  {Y.}~\bibnamefont {Wang}}, \bibinfo {author} {\bibfnamefont {V.~M.}\ \bibnamefont {Plisson}}, \bibinfo {author} {\bibfnamefont {M.}~\bibnamefont {Geiwitz}}, \bibinfo {author} {\bibfnamefont {C.~A.}\ \bibnamefont {Occhialini}}, \bibinfo {author} {\bibfnamefont {R.}~\bibnamefont {Comin}}, \bibinfo {author} {\bibfnamefont {M.~J.}\ \bibnamefont {Graf}}, \bibinfo {author} {\bibfnamefont {L.}~\bibnamefont {Zhao}}, \bibinfo {author} {\bibfnamefont {J.}~\bibnamefont {Cano}}, \bibinfo {author} {\bibfnamefont {R.~M.}\ \bibnamefont {Fernandes}}, \bibinfo {author} {\bibfnamefont {J.~J.}\ \bibnamefont {Cha}}, \bibinfo {author} {\bibfnamefont {L.~M.}\ \bibnamefont {Schoop}},\ and\ \bibinfo {author} {\bibfnamefont {K.~S.}\ \bibnamefont {Burch}},\ }\bibfield  {title} {\bibinfo {title} {Ferroaxial density wave from intertwined charge and orbital order in rare-earth tritellurides},\ }\href {https://doi.org/10.1038/s41567-025-03008-2} {\bibfield  {journal} {\bibinfo  {journal} {Nature Physics}\ }\textbf {\bibinfo {volume}
  {21}},\ \bibinfo {pages} {1578} (\bibinfo {year} {2025})}\BibitemShut {NoStop}%
\bibitem [{\citenamefont {Sacchetti}\ \emph {et~al.}(2006)\citenamefont {Sacchetti}, \citenamefont {Degiorgi}, \citenamefont {Giamarchi}, \citenamefont {Ru},\ and\ \citenamefont {Fisher}}]{Sacchetti2006}%
  \BibitemOpen
  \bibfield  {author} {\bibinfo {author} {\bibfnamefont {A.}~\bibnamefont {Sacchetti}}, \bibinfo {author} {\bibfnamefont {L.}~\bibnamefont {Degiorgi}}, \bibinfo {author} {\bibfnamefont {T.}~\bibnamefont {Giamarchi}}, \bibinfo {author} {\bibfnamefont {N.}~\bibnamefont {Ru}},\ and\ \bibinfo {author} {\bibfnamefont {I.~R.}\ \bibnamefont {Fisher}},\ }\bibfield  {title} {\bibinfo {title} {Chemical pressure and hidden one-dimensional behavior in rare-earth tri-telluride charge-density-wave compounds},\ }\href {https://doi.org/10.1103/PhysRevB.74.125115} {\bibfield  {journal} {\bibinfo  {journal} {Physical Review B}\ }\textbf {\bibinfo {volume} {74}},\ \bibinfo {pages} {125115} (\bibinfo {year} {2006})}\BibitemShut {NoStop}%
\bibitem [{\citenamefont {Pfuner}\ \emph {et~al.}(2009)\citenamefont {Pfuner}, \citenamefont {Degiorgi}, \citenamefont {Chu}, \citenamefont {Ru}, \citenamefont {Shin},\ and\ \citenamefont {Fisher}}]{Pfuner2008}%
  \BibitemOpen
  \bibfield  {author} {\bibinfo {author} {\bibfnamefont {F.}~\bibnamefont {Pfuner}}, \bibinfo {author} {\bibfnamefont {L.}~\bibnamefont {Degiorgi}}, \bibinfo {author} {\bibfnamefont {J.~H.}\ \bibnamefont {Chu}}, \bibinfo {author} {\bibfnamefont {N.}~\bibnamefont {Ru}}, \bibinfo {author} {\bibfnamefont {K.}~\bibnamefont {Shin}},\ and\ \bibinfo {author} {\bibfnamefont {I.}~\bibnamefont {Fisher}},\ }\bibfield  {title} {\bibinfo {title} {Optical properties of the charge-density-wave rare-earth tri-telluride compounds: A view on {PrTe$_3$}},\ }\href {https://doi.org/10.1016/j.physb.2008.11.052} {\bibfield  {journal} {\bibinfo  {journal} {Physica B: Condensed Matter}\ }\textbf {\bibinfo {volume} {404}},\ \bibinfo {pages} {533} (\bibinfo {year} {2009})}\BibitemShut {NoStop}%
\bibitem [{\citenamefont {Mennel}\ \emph {et~al.}(2018)\citenamefont {Mennel}, \citenamefont {Furchi}, \citenamefont {Wachter}, \citenamefont {Paur}, \citenamefont {Polyushkin},\ and\ \citenamefont {Mueller}}]{Mennel2018}%
  \BibitemOpen
  \bibfield  {author} {\bibinfo {author} {\bibfnamefont {L.}~\bibnamefont {Mennel}}, \bibinfo {author} {\bibfnamefont {M.~M.}\ \bibnamefont {Furchi}}, \bibinfo {author} {\bibfnamefont {S.}~\bibnamefont {Wachter}}, \bibinfo {author} {\bibfnamefont {M.}~\bibnamefont {Paur}}, \bibinfo {author} {\bibfnamefont {D.~K.}\ \bibnamefont {Polyushkin}},\ and\ \bibinfo {author} {\bibfnamefont {T.}~\bibnamefont {Mueller}},\ }\bibfield  {title} {\bibinfo {title} {Optical imaging of strain in two-dimensional crystals},\ }\href {https://doi.org/10.1038/s41467-018-02830-y} {\bibfield  {journal} {\bibinfo  {journal} {Nature Communications}\ }\textbf {\bibinfo {volume} {9}},\ \bibinfo {pages} {516} (\bibinfo {year} {2018})}\BibitemShut {NoStop}%
\bibitem [{\citenamefont {Narasimhamurty}(1981)}]{Narasimhamurty1981}%
  \BibitemOpen
  \bibfield  {author} {\bibinfo {author} {\bibfnamefont {T.~S.}\ \bibnamefont {Narasimhamurty}},\ }\href {https://doi.org/10.1007/978-1-4757-0025-1} {\emph {\bibinfo {title} {Photoelastic and Electro-Optic Properties of Crystals}}}\ (\bibinfo  {publisher} {Springer},\ \bibinfo {address} {Berlin, Heidelberg},\ \bibinfo {year} {1981})\BibitemShut {NoStop}%
\bibitem [{\citenamefont {Nye}(1957)}]{Nye1957}%
  \BibitemOpen
  \bibfield  {author} {\bibinfo {author} {\bibfnamefont {J.~F.}\ \bibnamefont {Nye}},\ }\href@noop {} {\emph {\bibinfo {title} {Physical Properties of Crystals: Their Representation by Tensors and Matrices}}}\ (\bibinfo  {publisher} {Oxford University Press},\ \bibinfo {address} {Oxford},\ \bibinfo {year} {1957})\BibitemShut {NoStop}%
\bibitem [{\citenamefont {Dav{\`\i}}(2020)}]{Davi2020}%
  \BibitemOpen
  \bibfield  {author} {\bibinfo {author} {\bibfnamefont {G.}~\bibnamefont {Dav{\`\i}}},\ }\bibfield  {title} {\bibinfo {title} {Exact and linearized refractive index stress-dependence in anisotropic photoelastic crystals},\ }\href {https://doi.org/10.1098/rspa.2019.0854} {\bibfield  {journal} {\bibinfo  {journal} {Proceedings of the Royal Society A}\ }\textbf {\bibinfo {volume} {476}},\ \bibinfo {pages} {20190854} (\bibinfo {year} {2020})}\BibitemShut {NoStop}%
\bibitem [{\citenamefont {Giannozzi}\ \emph {et~al.}(2009)\citenamefont {Giannozzi}, \citenamefont {Baroni}, \citenamefont {Bonini}, \citenamefont {Calandra}, \citenamefont {Car}, \citenamefont {Cavazzoni}, \citenamefont {Ceresoli}, \citenamefont {Chiarotti}, \citenamefont {Cococcioni}, \citenamefont {Dabo}, \citenamefont {Dal~Corso}, \citenamefont {de~Gironcoli}, \citenamefont {Fabris}, \citenamefont {Fratesi}, \citenamefont {Gebauer}, \citenamefont {Gerstmann}, \citenamefont {Gougoussis}, \citenamefont {Kokalj}, \citenamefont {Lazzeri}, \citenamefont {Martin-Samos}, \citenamefont {Marzari}, \citenamefont {Mauri}, \citenamefont {Mazzarello}, \citenamefont {Paolini}, \citenamefont {Pasquarello}, \citenamefont {Paulatto}, \citenamefont {Sbraccia}, \citenamefont {Scandolo}, \citenamefont {Sclauzero}, \citenamefont {Seitsonen}, \citenamefont {Smogunov}, \citenamefont {Umari},\ and\ \citenamefont {Wentzcovitch}}]{Giannozzi2009}%
  \BibitemOpen
  \bibfield  {author} {\bibinfo {author} {\bibfnamefont {P.}~\bibnamefont {Giannozzi}}, \bibinfo {author} {\bibfnamefont {S.}~\bibnamefont {Baroni}}, \bibinfo {author} {\bibfnamefont {N.}~\bibnamefont {Bonini}}, \bibinfo {author} {\bibfnamefont {M.}~\bibnamefont {Calandra}}, \bibinfo {author} {\bibfnamefont {R.}~\bibnamefont {Car}}, \bibinfo {author} {\bibfnamefont {C.}~\bibnamefont {Cavazzoni}}, \bibinfo {author} {\bibfnamefont {D.}~\bibnamefont {Ceresoli}}, \bibinfo {author} {\bibfnamefont {G.~L.}\ \bibnamefont {Chiarotti}}, \bibinfo {author} {\bibfnamefont {M.}~\bibnamefont {Cococcioni}}, \bibinfo {author} {\bibfnamefont {I.}~\bibnamefont {Dabo}}, \bibinfo {author} {\bibfnamefont {A.}~\bibnamefont {Dal~Corso}}, \bibinfo {author} {\bibfnamefont {S.}~\bibnamefont {de~Gironcoli}}, \bibinfo {author} {\bibfnamefont {S.}~\bibnamefont {Fabris}}, \bibinfo {author} {\bibfnamefont {G.}~\bibnamefont {Fratesi}}, \bibinfo {author} {\bibfnamefont {R.}~\bibnamefont {Gebauer}}, \bibinfo {author} {\bibfnamefont
  {U.}~\bibnamefont {Gerstmann}}, \bibinfo {author} {\bibfnamefont {C.}~\bibnamefont {Gougoussis}}, \bibinfo {author} {\bibfnamefont {A.}~\bibnamefont {Kokalj}}, \bibinfo {author} {\bibfnamefont {M.}~\bibnamefont {Lazzeri}}, \bibinfo {author} {\bibfnamefont {L.}~\bibnamefont {Martin-Samos}}, \bibinfo {author} {\bibfnamefont {N.}~\bibnamefont {Marzari}}, \bibinfo {author} {\bibfnamefont {F.}~\bibnamefont {Mauri}}, \bibinfo {author} {\bibfnamefont {R.}~\bibnamefont {Mazzarello}}, \bibinfo {author} {\bibfnamefont {S.}~\bibnamefont {Paolini}}, \bibinfo {author} {\bibfnamefont {A.}~\bibnamefont {Pasquarello}}, \bibinfo {author} {\bibfnamefont {L.}~\bibnamefont {Paulatto}}, \bibinfo {author} {\bibfnamefont {C.}~\bibnamefont {Sbraccia}}, \bibinfo {author} {\bibfnamefont {S.}~\bibnamefont {Scandolo}}, \bibinfo {author} {\bibfnamefont {G.}~\bibnamefont {Sclauzero}}, \bibinfo {author} {\bibfnamefont {A.~P.}\ \bibnamefont {Seitsonen}}, \bibinfo {author} {\bibfnamefont {A.}~\bibnamefont {Smogunov}}, \bibinfo {author}
  {\bibfnamefont {P.}~\bibnamefont {Umari}},\ and\ \bibinfo {author} {\bibfnamefont {R.~M.}\ \bibnamefont {Wentzcovitch}},\ }\bibfield  {title} {\bibinfo {title} {Quantum espresso: a modular and open-source software project for quantum simulations of materials},\ }\href {https://doi.org/10.1088/0953-8984/21/39/395502} {\bibfield  {journal} {\bibinfo  {journal} {Journal of Physics: Condensed Matter}\ }\textbf {\bibinfo {volume} {21}},\ \bibinfo {pages} {395502} (\bibinfo {year} {2009})}\BibitemShut {NoStop}%
\bibitem [{\citenamefont {Giannozzi}\ \emph {et~al.}(2017)\citenamefont {Giannozzi}, \citenamefont {Andreussi}, \citenamefont {Brumme}, \citenamefont {Bunau}, \citenamefont {Buongiorno~Nardelli}, \citenamefont {Calandra}, \citenamefont {Car}, \citenamefont {Cavazzoni}, \citenamefont {Ceresoli}, \citenamefont {Cococcioni}, \citenamefont {Colonna}, \citenamefont {Carnimeo}, \citenamefont {Dal~Corso}, \citenamefont {de~Gironcoli}, \citenamefont {Delugas}, \citenamefont {DiStasio}, \citenamefont {Ferretti}, \citenamefont {Floris}, \citenamefont {Fratesi}, \citenamefont {Fugallo}, \citenamefont {Gebauer}, \citenamefont {Gerstmann}, \citenamefont {Giustino}, \citenamefont {Gorni}, \citenamefont {Jia}, \citenamefont {Kawamura}, \citenamefont {Ko}, \citenamefont {Kokalj}, \citenamefont {Küçükbenli}, \citenamefont {Lazzeri}, \citenamefont {Marsili}, \citenamefont {Marzari}, \citenamefont {Mauri}, \citenamefont {Nguyen}, \citenamefont {Nguyen}, \citenamefont {Otero-de-la Roza}, \citenamefont {Paulatto},
  \citenamefont {Poncé}, \citenamefont {Rocca}, \citenamefont {Sabatini}, \citenamefont {Santra}, \citenamefont {Schlipf}, \citenamefont {Seitsonen}, \citenamefont {Smogunov}, \citenamefont {Timrov}, \citenamefont {Thonhauser}, \citenamefont {Umari}, \citenamefont {Vast}, \citenamefont {Wu},\ and\ \citenamefont {Baroni}}]{Giannozzi2017}%
  \BibitemOpen
  \bibfield  {author} {\bibinfo {author} {\bibfnamefont {P.}~\bibnamefont {Giannozzi}}, \bibinfo {author} {\bibfnamefont {O.}~\bibnamefont {Andreussi}}, \bibinfo {author} {\bibfnamefont {T.}~\bibnamefont {Brumme}}, \bibinfo {author} {\bibfnamefont {O.}~\bibnamefont {Bunau}}, \bibinfo {author} {\bibfnamefont {M.}~\bibnamefont {Buongiorno~Nardelli}}, \bibinfo {author} {\bibfnamefont {M.}~\bibnamefont {Calandra}}, \bibinfo {author} {\bibfnamefont {R.}~\bibnamefont {Car}}, \bibinfo {author} {\bibfnamefont {C.}~\bibnamefont {Cavazzoni}}, \bibinfo {author} {\bibfnamefont {D.}~\bibnamefont {Ceresoli}}, \bibinfo {author} {\bibfnamefont {M.}~\bibnamefont {Cococcioni}}, \bibinfo {author} {\bibfnamefont {N.}~\bibnamefont {Colonna}}, \bibinfo {author} {\bibfnamefont {I.}~\bibnamefont {Carnimeo}}, \bibinfo {author} {\bibfnamefont {A.}~\bibnamefont {Dal~Corso}}, \bibinfo {author} {\bibfnamefont {S.}~\bibnamefont {de~Gironcoli}}, \bibinfo {author} {\bibfnamefont {P.}~\bibnamefont {Delugas}}, \bibinfo {author} {\bibfnamefont
  {R.~A.}\ \bibnamefont {DiStasio}}, \bibinfo {author} {\bibfnamefont {A.}~\bibnamefont {Ferretti}}, \bibinfo {author} {\bibfnamefont {A.}~\bibnamefont {Floris}}, \bibinfo {author} {\bibfnamefont {G.}~\bibnamefont {Fratesi}}, \bibinfo {author} {\bibfnamefont {G.}~\bibnamefont {Fugallo}}, \bibinfo {author} {\bibfnamefont {R.}~\bibnamefont {Gebauer}}, \bibinfo {author} {\bibfnamefont {U.}~\bibnamefont {Gerstmann}}, \bibinfo {author} {\bibfnamefont {F.}~\bibnamefont {Giustino}}, \bibinfo {author} {\bibfnamefont {T.}~\bibnamefont {Gorni}}, \bibinfo {author} {\bibfnamefont {J.}~\bibnamefont {Jia}}, \bibinfo {author} {\bibfnamefont {M.}~\bibnamefont {Kawamura}}, \bibinfo {author} {\bibfnamefont {H.-Y.}\ \bibnamefont {Ko}}, \bibinfo {author} {\bibfnamefont {A.}~\bibnamefont {Kokalj}}, \bibinfo {author} {\bibfnamefont {E.}~\bibnamefont {Küçükbenli}}, \bibinfo {author} {\bibfnamefont {M.}~\bibnamefont {Lazzeri}}, \bibinfo {author} {\bibfnamefont {M.}~\bibnamefont {Marsili}}, \bibinfo {author} {\bibfnamefont
  {N.}~\bibnamefont {Marzari}}, \bibinfo {author} {\bibfnamefont {F.}~\bibnamefont {Mauri}}, \bibinfo {author} {\bibfnamefont {N.~L.}\ \bibnamefont {Nguyen}}, \bibinfo {author} {\bibfnamefont {H.-V.}\ \bibnamefont {Nguyen}}, \bibinfo {author} {\bibfnamefont {A.}~\bibnamefont {Otero-de-la Roza}}, \bibinfo {author} {\bibfnamefont {L.}~\bibnamefont {Paulatto}}, \bibinfo {author} {\bibfnamefont {S.}~\bibnamefont {Poncé}}, \bibinfo {author} {\bibfnamefont {D.}~\bibnamefont {Rocca}}, \bibinfo {author} {\bibfnamefont {R.}~\bibnamefont {Sabatini}}, \bibinfo {author} {\bibfnamefont {B.}~\bibnamefont {Santra}}, \bibinfo {author} {\bibfnamefont {M.}~\bibnamefont {Schlipf}}, \bibinfo {author} {\bibfnamefont {A.~P.}\ \bibnamefont {Seitsonen}}, \bibinfo {author} {\bibfnamefont {A.}~\bibnamefont {Smogunov}}, \bibinfo {author} {\bibfnamefont {I.}~\bibnamefont {Timrov}}, \bibinfo {author} {\bibfnamefont {T.}~\bibnamefont {Thonhauser}}, \bibinfo {author} {\bibfnamefont {P.}~\bibnamefont {Umari}}, \bibinfo {author}
  {\bibfnamefont {N.}~\bibnamefont {Vast}}, \bibinfo {author} {\bibfnamefont {X.}~\bibnamefont {Wu}},\ and\ \bibinfo {author} {\bibfnamefont {S.}~\bibnamefont {Baroni}},\ }\bibfield  {title} {\bibinfo {title} {Advanced capabilities for materials modelling with quantum espresso},\ }\href {https://doi.org/10.1088/1361-648x/aa8f79} {\bibfield  {journal} {\bibinfo  {journal} {Journal of Physics: Condensed Matter}\ }\textbf {\bibinfo {volume} {29}},\ \bibinfo {pages} {465901} (\bibinfo {year} {2017})}\BibitemShut {NoStop}%
\bibitem [{\citenamefont {Jain}\ \emph {et~al.}(2013)\citenamefont {Jain}, \citenamefont {Ong}, \citenamefont {Hautier}, \citenamefont {Chen}, \citenamefont {Richards}, \citenamefont {Dacek}, \citenamefont {Cholia}, \citenamefont {Gunter}, \citenamefont {Skinner}, \citenamefont {Ceder},\ and\ \citenamefont {Persson}}]{Jain2013}%
  \BibitemOpen
  \bibfield  {author} {\bibinfo {author} {\bibfnamefont {A.}~\bibnamefont {Jain}}, \bibinfo {author} {\bibfnamefont {S.~P.}\ \bibnamefont {Ong}}, \bibinfo {author} {\bibfnamefont {G.}~\bibnamefont {Hautier}}, \bibinfo {author} {\bibfnamefont {W.}~\bibnamefont {Chen}}, \bibinfo {author} {\bibfnamefont {W.~D.}\ \bibnamefont {Richards}}, \bibinfo {author} {\bibfnamefont {S.}~\bibnamefont {Dacek}}, \bibinfo {author} {\bibfnamefont {S.}~\bibnamefont {Cholia}}, \bibinfo {author} {\bibfnamefont {D.}~\bibnamefont {Gunter}}, \bibinfo {author} {\bibfnamefont {D.}~\bibnamefont {Skinner}}, \bibinfo {author} {\bibfnamefont {G.}~\bibnamefont {Ceder}},\ and\ \bibinfo {author} {\bibfnamefont {K.~A.}\ \bibnamefont {Persson}},\ }\bibfield  {title} {\bibinfo {title} {Commentary: The {Materials Project}: A materials genome approach to accelerating materials innovation},\ }\href {https://doi.org/10.1063/1.4812323} {\bibfield  {journal} {\bibinfo  {journal} {APL Materials}\ }\textbf {\bibinfo {volume} {1}},\ \bibinfo {pages}
  {011002} (\bibinfo {year} {2013})}\BibitemShut {NoStop}%
\bibitem [{\citenamefont {Ong}\ \emph {et~al.}(2013)\citenamefont {Ong}, \citenamefont {Richards}, \citenamefont {Jain}, \citenamefont {Hautier}, \citenamefont {Kocher}, \citenamefont {Cholia}, \citenamefont {Gunter}, \citenamefont {Chevrier}, \citenamefont {Persson},\ and\ \citenamefont {Ceder}}]{Ong2013}%
  \BibitemOpen
  \bibfield  {author} {\bibinfo {author} {\bibfnamefont {S.~P.}\ \bibnamefont {Ong}}, \bibinfo {author} {\bibfnamefont {W.~D.}\ \bibnamefont {Richards}}, \bibinfo {author} {\bibfnamefont {A.}~\bibnamefont {Jain}}, \bibinfo {author} {\bibfnamefont {G.}~\bibnamefont {Hautier}}, \bibinfo {author} {\bibfnamefont {M.}~\bibnamefont {Kocher}}, \bibinfo {author} {\bibfnamefont {S.}~\bibnamefont {Cholia}}, \bibinfo {author} {\bibfnamefont {D.}~\bibnamefont {Gunter}}, \bibinfo {author} {\bibfnamefont {V.~L.}\ \bibnamefont {Chevrier}}, \bibinfo {author} {\bibfnamefont {K.~A.}\ \bibnamefont {Persson}},\ and\ \bibinfo {author} {\bibfnamefont {G.}~\bibnamefont {Ceder}},\ }\bibfield  {title} {\bibinfo {title} {{Python Materials Genomics} (pymatgen): A robust, open-source {Python} library for materials analysis},\ }\href {https://doi.org/10.1016/j.commatsci.2012.10.028} {\bibfield  {journal} {\bibinfo  {journal} {Computational Materials Science}\ }\textbf {\bibinfo {volume} {68}},\ \bibinfo {pages} {314} (\bibinfo {year}
  {2013})}\BibitemShut {NoStop}%
\bibitem [{\citenamefont {Persson}(2016)}]{mposti_1313190}%
  \BibitemOpen
  \bibfield  {author} {\bibinfo {author} {\bibfnamefont {K.}~\bibnamefont {Persson}},\ }\href {https://doi.org/10.17188/1313190} {\bibinfo {title} {Materials data on {SmTe$_3$ (SG:63)} by {Materials Project}}} (\bibinfo {year} {2016}),\ \bibinfo {note} {an optional note}\BibitemShut {NoStop}%
\bibitem [{\citenamefont {Stoehr}\ \emph {et~al.}(2020)\citenamefont {Stoehr}, \citenamefont {Gerlach}, \citenamefont {H\"artling},\ and\ \citenamefont {Schoenfelder}}]{Stoehr2020}%
  \BibitemOpen
  \bibfield  {author} {\bibinfo {author} {\bibfnamefont {M.}~\bibnamefont {Stoehr}}, \bibinfo {author} {\bibfnamefont {G.}~\bibnamefont {Gerlach}}, \bibinfo {author} {\bibfnamefont {T.}~\bibnamefont {H\"artling}},\ and\ \bibinfo {author} {\bibfnamefont {S.}~\bibnamefont {Schoenfelder}},\ }\bibfield  {title} {\bibinfo {title} {Analysis of photoelastic properties of monocrystalline silicon},\ }\href {https://doi.org/10.5194/jsss-9-209-2020} {\bibfield  {journal} {\bibinfo  {journal} {Journal of Sensors and Sensor Systems}\ }\textbf {\bibinfo {volume} {9}},\ \bibinfo {pages} {209} (\bibinfo {year} {2020})}\BibitemShut {NoStop}%
\bibitem [{\citenamefont {Singh}\ \emph {et~al.}(2024)\citenamefont {Singh}, \citenamefont {Bachmann}, \citenamefont {Sanchez}, \citenamefont {Pandey}, \citenamefont {Kapitulnik}, \citenamefont {Kim}, \citenamefont {Ryan}, \citenamefont {Kivelson},\ and\ \citenamefont {Fisher}}]{singh2024}%
  \BibitemOpen
  \bibfield  {author} {\bibinfo {author} {\bibfnamefont {A.~G.}\ \bibnamefont {Singh}}, \bibinfo {author} {\bibfnamefont {M.~D.}\ \bibnamefont {Bachmann}}, \bibinfo {author} {\bibfnamefont {J.~J.}\ \bibnamefont {Sanchez}}, \bibinfo {author} {\bibfnamefont {A.}~\bibnamefont {Pandey}}, \bibinfo {author} {\bibfnamefont {A.}~\bibnamefont {Kapitulnik}}, \bibinfo {author} {\bibfnamefont {J.~W.}\ \bibnamefont {Kim}}, \bibinfo {author} {\bibfnamefont {P.~J.}\ \bibnamefont {Ryan}}, \bibinfo {author} {\bibfnamefont {S.~A.}\ \bibnamefont {Kivelson}},\ and\ \bibinfo {author} {\bibfnamefont {I.~R.}\ \bibnamefont {Fisher}},\ }\bibfield  {title} {\bibinfo {title} {Emergent tetragonality in a fundamentally orthorhombic material},\ }\href {https://doi.org/10.1126/sciadv.adk3321} {\bibfield  {journal} {\bibinfo  {journal} {Science Advances}\ }\textbf {\bibinfo {volume} {10}},\ \bibinfo {pages} {eadk3321} (\bibinfo {year} {2024})}\BibitemShut {NoStop}%
\bibitem [{\citenamefont {Powell}(2010)}]{Powell2010}%
  \BibitemOpen
  \bibfield  {author} {\bibinfo {author} {\bibfnamefont {R.~C.}\ \bibnamefont {Powell}},\ }\href {https://doi.org/10.1007/978-1-4419-7598-0} {\emph {\bibinfo {title} {Symmetry, Group Theory, and the Physical Properties of Crystals}}},\ \bibinfo {series} {Lecture Notes in Physics}, Vol.\ \bibinfo {volume} {824}\ (\bibinfo  {publisher} {Springer},\ \bibinfo {year} {2010})\BibitemShut {NoStop}%
\end{thebibliography}%

\end{document}